\documentclass[floatfix, noshowpacs, preprintnumbers, twocolumn, amsmath, amssymb, aps, prl, superscriptaddress]{revtex4-1}

\pdfoutput=1

\usepackage{graphicx}
\usepackage{amsmath,soul,tabularx,mathtools}
\usepackage{amssymb}
\usepackage{times}
\usepackage{bbm}
\usepackage{bm}
\usepackage[T1]{fontenc}
\usepackage{bbold}
\usepackage{nccmath}
\usepackage[mathscr]{eucal}
\usepackage{multirow}
\usepackage{colortbl}
\usepackage[utf8]{inputenc}
\usepackage{lmodern}
\usepackage{xcolor}
\usepackage[caption=false]{subfig}
\usepackage{ifthen}

\begin{document}

\title{Benchmarking integrated photonic architectures}

\author{Fulvio Flamini}
\affiliation{Dipartimento di Fisica, Sapienza Universit\`{a} di Roma,
Piazzale Aldo Moro 5, I-00185 Roma, Italy}

\author{Nicol\`o Spagnolo}
\affiliation{Dipartimento di Fisica, Sapienza Universit\`{a} di Roma,
Piazzale Aldo Moro 5, I-00185 Roma, Italy}

\author{Niko Viggianiello}
\affiliation{Dipartimento di Fisica, Sapienza Universit\`{a} di Roma,
Piazzale Aldo Moro 5, I-00185 Roma, Italy}

\author{Andrea Crespi}
\affiliation{Istituto di Fotonica e Nanotecnologie, Consiglio Nazionale delle Ricerche (IFN-CNR), 
Piazza Leonardo da Vinci, 32, I-20133 Milano, Italy}
\affiliation{Dipartimento di Fisica, Politecnico di Milano, Piazza Leonardo da Vinci, 32, I-20133 Milano, Italy}

\author{Roberto Osellame}
\affiliation{Istituto di Fotonica e Nanotecnologie, Consiglio Nazionale delle Ricerche (IFN-CNR), 
Piazza Leonardo da Vinci, 32, I-20133 Milano, Italy}
\affiliation{Dipartimento di Fisica, Politecnico di Milano, Piazza Leonardo da Vinci, 32, I-20133 Milano, Italy}

\author{Fabio Sciarrino}
\affiliation{Dipartimento di Fisica, Sapienza Universit\`{a} di Roma,
Piazzale Aldo Moro 5, I-00185 Roma, Italy}

\begin{abstract}
Photonic platforms represent a promising technology for the realization of several quantum communication protocols and for experiments of quantum simulation. Moreover, large-scale integrated interferometers have recently gained a relevant role for restricted models of quantum computing, specifically with Boson Sampling devices. Indeed, various linear optical schemes have been proposed for the implementation of unitary transformations, each one suitable for a specific task. Notwithstanding, so far a comprehensive analysis of the state of the art under broader and realistic conditions is still lacking.
In the present work we address this gap, providing in a unified framework a quantitative comparison of the three main photonic architectures, namely the ones with triangular and square designs and the so-called fast transformations. All layouts have been analyzed in presence of losses and imperfect control over the reflectivities and phases of the inner structure. Our results represent a further step ahead towards the implementation of quantum information protocols on large-scale integrated photonic devices.
\end{abstract}

\maketitle

\begin{figure*}[ht!]
\centering
\includegraphics[width=0.99\textwidth]{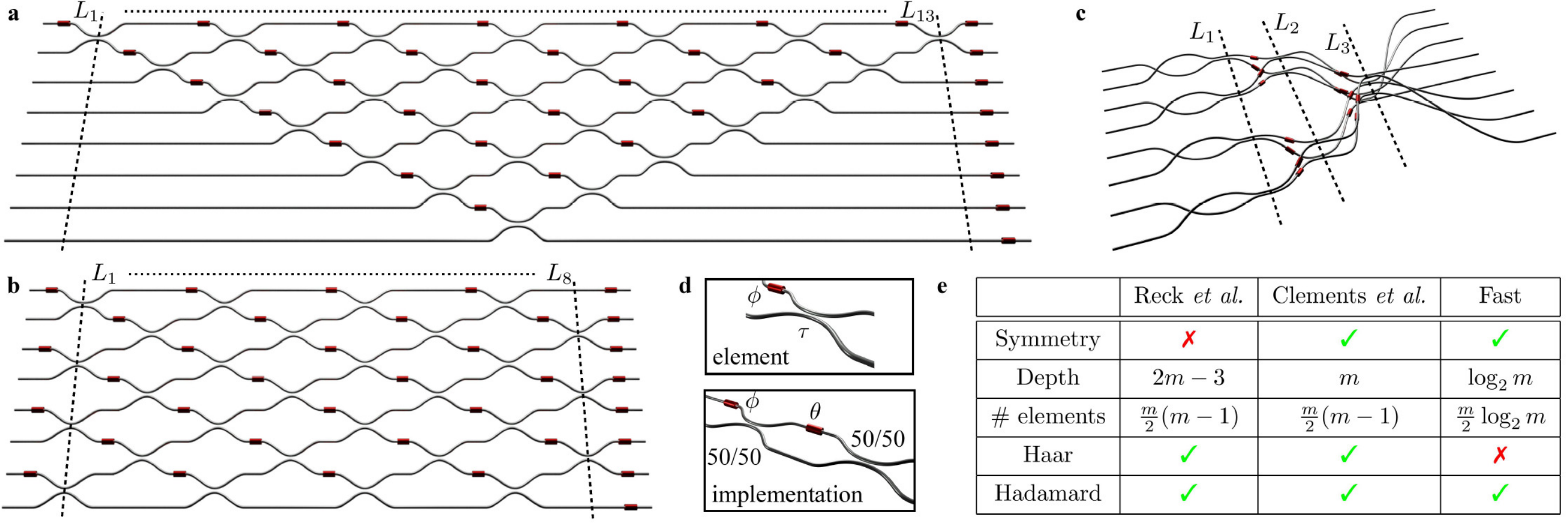}
\caption{{\bf Architectures for integrated photonic networks}. Interferometric layouts for a $8$-mode unitary transformation with ({\bf a}) the triangular scheme of Reck {\em et al.} \cite{Reck} and with ({\bf b}) the scheme of Clements {\em et al.} \cite{Clements16}. For both architectures, unit cells consist of one beam splitter with arbitrary transmittivity $\tau$ and one phase shift $\varphi$ (red cylinders). {\bf c}, Fast architecture with 3-dimensional layout shown for a $8$-mode interferometer, which realizes a significant class of transformations with a reduced number of layers and optical elements. {\bf d}, Unit cell for the three layouts. Beam splitters in each unit cell can be implemented as a Mach-Zehnder interferometer composed by two symmetric (50/50) beam splitters and an internal relative phase shift $\theta$. {\bf e}, Table summarizing the main features of the three architectures.}
\label{fig:schemi}
\end{figure*}

Several milestone achievements in experimental quantum information are pushing the limits of integrated photonic technologies in numerous relevant applications. Single-photon sources \cite{Silverstone2014,Somaschi2016,Spring2017,Montaut2017} and detectors \cite{Lita2008,Calkins2013} are already providing remarkable results in first benchmark demonstrations, while a number of powerful techniques have been developed to fully characterize general quantum processes \cite{OBrien2004,Lobino2008,Laing12,Rahimi2013,Tillmann2016,Spagnolo2016}. The miniaturization of complex interferometric schemes is thus expected to unlock stable and mass-produced large-scale quantum information protocols, among the others for teleportation \cite{Metcalf14}, logic gates \cite{Politi2009,Crespi2011}, quantum networks \cite{Peruzzo10,Sansoni12} and light manipulation \cite{Tanzilli12,Orieux16}. One further, fundamental feature of such platforms is the capability of adding dynamical reconfigurability to the circuits \cite{Smith09,Matthews09,Chaboyer15,Flamini15}, allowing for universal applications as for standard classical processors \cite{Carolan15,Harris15,Miller15}. More in particular, a research area that well benefits from all the above-mentioned technologies is that of Boson Sampling \cite{Broome13,Spring13,Spagnolo13prl,Tillmann13,Crespi13,Spagnolo14,Carolan14,Bentivegna15,Aaronson2011}, where efficient sampling from linear photonic devices could provide evidence of a quantum computational power beyond the reach of classical computers \cite{Aaronson2011,Aaronson15,Latmiral16}.\\
In this context, it is essential to identify suitable architectures to implement large-size interferometric networks within an integrated platform. In \cite{Reck}, Reck {\em et al.} proposed a universal algorithm to implement an arbitrary unitary transformation by decomposing it in a suitable network of unit cells made up of only beam splitters and phase shifters. For each transformation to be implemented, it is sufficient to determine the correct set of parameters (beam splitter transmissivities and internal phases) without altering the overall interferometric layout. This architecture has been employed in first experimental instances of Boson Sampling \cite{Crespi13}, where the capability to implement arbitrary Haar random unitaries is an essential ingredient for the demonstration of its computational complexity. However, this architecture lacks a perfect symmetry in its triangular layout, thus making it sensitive to internal losses that lower the adherence of the implemented transformation to the ideal one. Recently, two different architectures have been proposed for the implementation of linear optical networks. A first scheme has been reported in \cite{Clements16}, which ultimately corresponds to a cunning rearrangement of the scheme of \cite{Reck} in a symmetric layout. While keeping the same number of optical elements and the capability of implementing an arbitrary unitary transformation, this scheme presents reduced sensitivity to losses within the interferometer. A second scheme, inspired by the classical algorithm of Cooley and Tukey \cite{Cooley65} for the fast Fourier transform, has been proposed recently in \cite{Torma96,Barak07} and implemented experimentally in \cite{Crespi2016,Flamini16, Viggianiello17} by exploiting the three-dimensional capabilities of femtosecond laser micromachining \cite{gattass2008flm,Spagnolo13}. This layout, though not supporting arbitrary unitary evolutions, allows to implement a significant class of linear optical networks with a substantial reduction in the number of necessary optical elements. Such class of matrices includes the Hadamard ones, with a notable example provided by the Fourier transformation that is widely employed in a large set of quantum information protocols \cite{Shor94,Nielsen00}. A crucial requirement towards the identification of optimal architectures for the implementation of large size interferometers is a detailed knowledge of the tolerance to fabricative errors, namely propagation losses and imperfections in the parameters of the optical elements. Indeed, in all quantum information applications including Boson Sampling \cite{Arkhipov14,Leverrier15} the applicability of an experimental platform is limited by the maximum amount of noise tolerable in the interfometric networks. Within this framework, a thorough analysis of the tolerance of the proposed architectures in the presence of fabrication noise is still lacking.

In this article we address this gap, presenting a complete analysis of the performance of the main photonic architectures under imperfect operational conditions. The three interferometric layouts have been investigated in the general case, by admitting different levels of losses and noise in unitary transformations of increasing size. Specifically, following the state-of-the-art approach adopted for reconfigurable quantum circuits \cite{Carolan15,Harris15}, i.e. by modelling beam splitters as Mach-Zehnder interferometers with variable phases and two cascaded symmetric beam splitters, noise was added to their reflectivities and to phases in both Mach-Zehnders and outer phase shifters.
For our numerical benchmark we employ as figures of merit the fidelity \cite{Clements16} and the total variation distance, as good estimators of the distance between ideal and imperfect implementations in relevant applications. The article is structured as follows. First, we briefly discuss the interferometric structure of the triangular, square and fast designs. Thereafter, we compare the performances of the first two schemes, which were shown to be universal for unitary evolutions, for the implementation of Haar-random transformations of increasing size. Finally, we compare the operation of the three schemes for the implementation of Fourier and Sylvester interferometers, which represent the fundamental building blocks in a significant number of relevant quantum information protocols. Our analysis highlights the advantages and limitations of each scheme, providing a reference point for the design of future larger-scale photonic technologies, whose optimal configurations may well benefit from a joint integration.

\section*{Results}

\subsection*{Designs for photonic architectures} 

Since the seminal work of Hurwitz \cite{Hurwitz97}, it is known that every $m\times m$ unitary transformation can be decomposed in the action of $\frac{m(m-1)}{2} $ unitaries, each acting on a two-dimensional subspace of the Hilbert space. Reck \textit{et al.} \cite{Reck} independently gave the first operational proof ($R$) that an actual (linear optical) implementation, consisting of only single-mode phase shifters and two-mode beam splitters, does exist for any discrete unitary operator. Recently, a new algorithm ($C$) for the decomposition of arbitrary unitary transformations has been introduced by Clements \textit{et al.} \cite{Clements16}, which basically presents a higher resilience against propagation losses thanks to the compact and fully symmetric design.
Both $R$ and $C$ decompositions are made up of a set of $N=\frac{m(m-1)}{2}$ unitaries $T_{k,k+1}^{(i)}$, each coupling step by step modes $k$ and $k+1$ of the interferometer

\begin{equation}
	T_{k,k+1}^{(i)} =
	\left(
	\begin{array}{cccc}
		\mathbb{1} &  &  & 0 \\
		 & - \sin \omega_i & e^{- \imath \phi_i} \, \cos \omega_i   &  \\
		 & \cos \omega_i & e^{- \imath \phi_i} \, \sin \omega_i  &  \\
		0 &  &  & \mathbb{1}
	\end{array} \right)
\label{eq:decompositionRC}
\end{equation}

\noindent where $U_{R,C} = \prod_{i=1}^N T_{k,k+1}^{(i)}$ and the order of the interactions is directly related to the triangular and square designs of, respectively, the $R$ and $C$ schemes.
While being both universal for the decomposition of unitary transformations, the $C$-design presents some immediate advantages in terms of circuit depth, namely a more balanced mixing of the optical modes and less propagation losses thanks to a minimized operation area, which is also a crucial requirement for large-scale implementations.

\begin{figure*}[t!]
\centering
\includegraphics[width=0.99\textwidth]{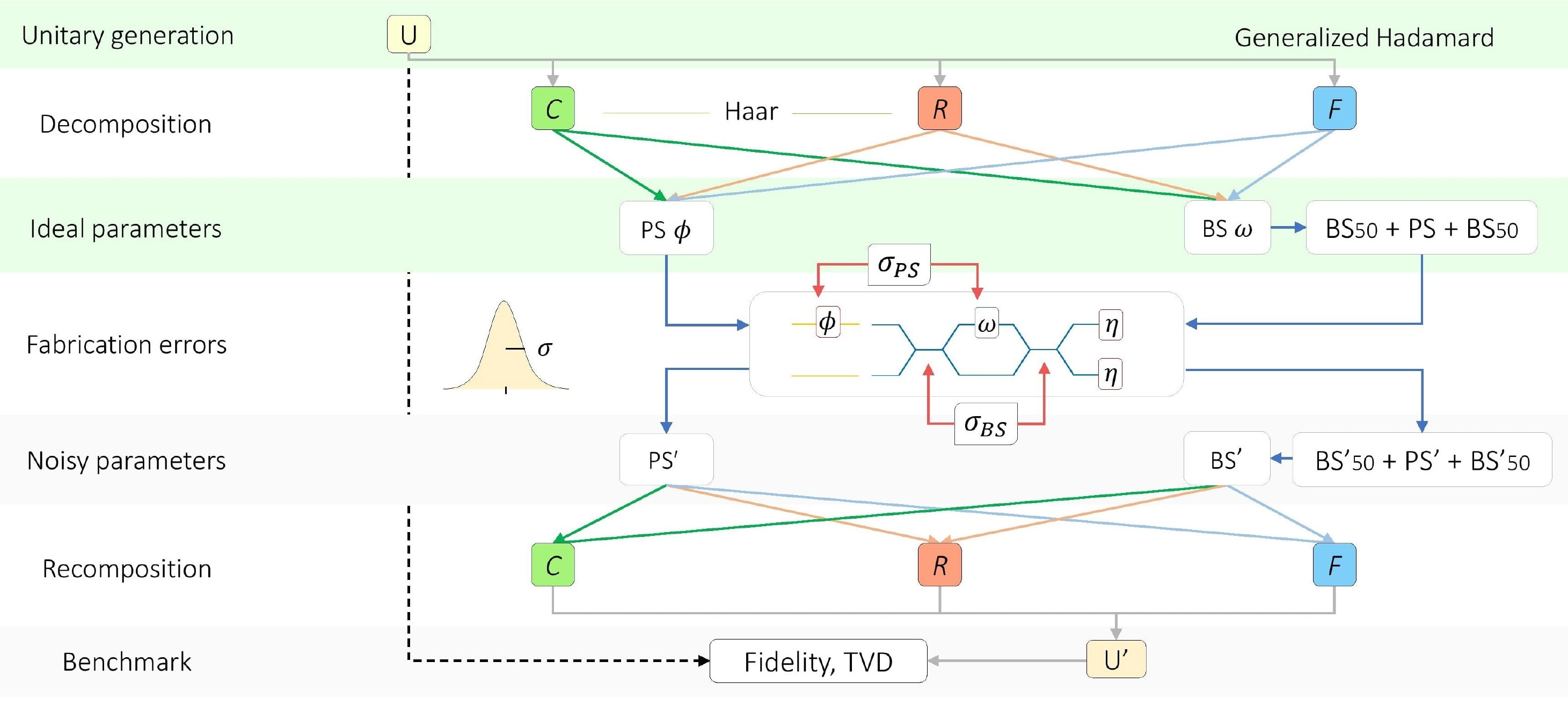}
\caption{{\bf Benchmarking imperfect implementations}. Performance of photonic architectures ($C$: Clements \cite{Clements16}, in green; $R$: Reck \cite{Reck}, in red, $F$: Fast, in blue) is investigated under realistic conditions of noise and losses. Throughout our analysis, simulations are carried out by introducing gaussian noise separately on both phase shifters ($\sigma_{PS}$) and beam splitters ($\sigma_{BS}$) in their Mach-Zehnder implementation. As figure of merit for their performance we adopt the fidelity \cite{Clements16} and the total variation distance (TVD) to address more general applications involving also multiphoton evolutions. }
\label{fig:analysis}
\end{figure*}

The universality of the $C$- and $R$-designs for unitary decomposition comes however at the cost of ignoring possible symmetries of the transformations, which could reduce the complexity of specific implementations. A relevant example is represented for instance by the Hadamard transformations, whose symmetries are known to allow remarkable simplifications in their algorithmic construction \cite{Cooley65}. 
Notable representatives of the Hadamard class are the Fourier ($U^{\operatorname{F}}$) and Sylvester ($U^{\operatorname{S}}$) transformations, respectively described by the $2^n$-dimensional unitaries $U_{a,b}^{\operatorname{F}}(2^n)=\frac{1}{\sqrt{2^n}}\, e^{2 \pi   i \frac{a b}{2^n}}$ and

\begin{equation}
	U^{\operatorname{S}}(2^n)=S(2^n)=\left(
		\begin{array}{cc}
			S(2^{n-1}) & S(2^{n-1}) \\
			S(2^{n-1}) & -S(2^{n-1})
		\end{array} \right)
\end{equation}

being $S(2^0)=(1)$ and $n$ any positive integer.
More generally, starting from the linear-optical fast Fourier decomposition developed by Barak \textit{et al.} \cite{Barak07}, it is possible to generalize their scheme to span a whole class of generalized Hadamard transformations \cite{Flamini17} by keeping fixed the interferometric structure and tuning the parameters of the beam splitters and phase shifters. The essential structure of such fast architectures presents some interesting advantages with respect to the other universal schemes. First, the depth of the circuit scales only logarithmically with the size of the interferometer, i.e. with the number of optical modes, leading to an even more compact operation area and to reduced propagation losses. Moreover, the layout is fully symmetric and naturally fits a description in terms of qubit states, thanks to the binary interactions between the modes.
A closed-form expression of the element $ U_{a,b} $ of the most general $2^n$-dimensional fast unitary transformation has the form

\begin{equation}
U_{\,a+1,\,b+1}^{(n)}= e^{i \, b_r \; \phi_r^{\xi_r^{(a,b)}} + \, i \, \frac{\pi}{2} \, a_r^{(n)} \oplus b_r^{(n)} }\; \prod_{s=1}^n \cos \chi_{a,b}^{(n,s)}
\label{formula}
\end{equation}

\noindent where $a,b \in \left[0,2^n-1\right]$ label the input/output modes and some shorthand notations have been used, following the Einstein summation convention

\begin{equation}
\begin{array}{c}
\chi_{a,b}^{(n,s)} = \theta_{s,f_s^{(a,b)}} - \frac{\pi}{2} \,| a_s^{(n)}-b_s^{(n)} | \\ \\
 \xi_r^{(a,b)} = 1+ \alpha_r + b \;\textup{mod}  \,\alpha_r + \lfloor \, \frac{a}{2 \alpha_r} \, \rfloor \, 2\alpha_r
\end{array}
\end{equation}

\noindent where $\alpha_r=2^{n-r}$, $ f_s^{(a,b)} =1+b+ ( a_r^{(n)}-b_r^{(n)} ) \, 2^{n-r} $ and $ m_r^{(n)} $ equals the $r$-th digit of the $n$-bit binary representation of $m$, being $ \lfloor x \rfloor $ the integer part of $x$.

In general, any $m$-dimensional photonic architectures can be described in terms of consecutive layers $s$ of optical elements $L_s$, made up of a network of phase shifters and beam splitters mixing a subset of modes no more than once each.
Specifically, each matrix $L_s$ consists in turn of a layer $B^{(s)}$ of beam splitters, coupling a set of pairs of modes $(k_1, k_2)$, and $\frac{m}{2}$ phase shifters $e^{i \phi_{s,k_1}}$ placed for each pair on one of the two interacting modes. The particular sequence of mode interactions $\{(k_1, k_2)\}$ depends on the triangular, square or fast designs. While clearer for the first two, the geometry of the third scheme for a $2^n$-dimensional unitary transformation is slightly more complex and arises from the binary representations of the optical modes. Using $\tau_{s,k}$ for the beam splitters transmissivities on mode $k$, the beamsplitters layer is described by the matrix

\begin{equation}
B_{k_1,k_2}^{(s)}\equiv
\left\{
                \begin{array}{ll}
                  \ \tau_{s,k_{1}} \hspace{7em}  k_1=k_2 \\
                  \ i  \sqrt{1-\tau_{s,k_1}^{2}}  \,\quad (k_1,k_2) \in \{(\alpha, \beta)\}^{(n,s)} \\
\ \\
                  \ 0  \hspace{8em} \textit{otherwise}\\
                \end{array}
                           \right.            
\label{eq:layerBS}                        
\end{equation}

\noindent where $\{(\alpha, \beta) \}^{(n,s)}=\{ (\: a+2^{s}\: b,\: a+2^{s}\: b+2^{s-1}\, ) \}$ are the pairs of modes interacting in the layer $s$, with $ a \in {\{1,..,2^{s-1} \}}\,,\; b \in { \{ 0,..,2^{n-s}-1\}}$. For example, the 8-dimensional quantum Fourier transform is obtained, modulo a relabeling of the output modes \cite{Barak07}, by choosing $ \tau_{s,k}=\sqrt{2^{-1}} $  and $\phi_{2,7}=\phi_{2,8}=\phi_{3,4}=\frac{\pi}{2}$,  $\phi_{3,6}=\frac{\pi}{4}$ and $\phi_{3,8}=\frac{3 \pi}{4}$. Similarly, the Sylvester transformation corresponds to the choice $ \tau_{s,k}=\sqrt{2^{-1}} $  and $\phi_{s,k}=0$.

\subsection*{Modelling non-ideal unitary implementations} 

We can now introduce the model adopted to probe the three architectures under non-ideal conditions. In the following we will refer to the $C$- (Clements \textit{et al.} \cite{Clements16}), $R-$  (Reck \textit{et al.} \cite{Reck}) and $F$-  (Fast) designs looking at their fixed, solid photonic architectures. This aspect is especially relevant if we are to choose the layout of a fully reconfigurable quantum circuit, which is designed to be multi-purpose and optimal when averaging over all its applications of interest. In general, the two main factors affecting the implementation of photonic quantum circuits are propagation losses and imperfect settings of the parameters describing the optical elements.

\textit{Losses --} Coupling losses at the input/output of any circuits remain relevant aspects in practical situations; however, their effect can be regarded as independent of the internal photonic architecture adopted and, thus, they will not be included in our study. Propagation losses, occurring in both straight and bent waveguides, play instead the main role in spoiling multipath interference. Their impact was shown to be mitigated by a more compact and symmetric interferometric structure \cite{Clements16}, where optical modes interact with each other in a balanced way. Though photon losses unavoidably occurr all along the circuit, a simple and effective way to analyze their effect is that of inserting costant losses at the output of each two-mode unit cell.

\textit{Fabricative noise --} Imperfect control over the fabricative parameters naturally leads to deviations from the ideal evolution on the circuit while keeping the unitarity of the process. The source of noise, arising from imperfect fabrication of the beam splitters and phase shifters associated to the $T_{k,k+1}^{(i)}$ in Eq.(\ref{eq:decompositionRC}), has a different nature depending on the practical realization. Recent technological achievements have enabled the realization of reconfigurable quantum circuits \cite{Carolan15,Harris15}, where generic beam splitters are implemented as Mach-Zehnder interferometers with two cascaded symmetric beam splitters and one tunable phase shift. In this case, which will be at the core of our analysis, noise arises from imperfect fabrication of the fixed symmetric beam splitters and from a non-perfect control over the thermo-electric or electro-optic phase manipulation, which becomes non-negligible for large-scale circuits or with high-speed tunings.

\vspace{1em}

Following the scheme of Fig.\ref{fig:analysis}, our investigation on the performances of the three architectures was carried out by introducing various levels of noise on the optical elements and losses after each unit cell. Our Monte Carlo simulations proceed through the following steps:
\begin{enumerate}
	\item sample a unitary transformation $U$ according to the Haar measure;
	\item apply $C$ or $R$ algorithms to retrieve the parameters $(\omega_i,\phi_i)$ according to Eq.(\ref{eq:decompositionRC});
	\item implement each $T_{k,k+1}^{(i)}$ as a Mach-Zehnder with input phase shift $\phi_i$ and $\tau_{i,1}=\tau_{i,2}=2^{-1/2}$:
	\begin{widetext}
	\begin{equation}
			T^{(i)} \: \rightarrow \:
			- \imath
			\left( \begin{array}{cc}
				 \tau_{i,2} & \imath \sqrt{1-\tau_{i,2}^2}  \\
				 \imath \sqrt{1-\tau_{i,2}^2} & \tau_{i,2}  \\
			\end{array} \right)
			\left( \begin{array}{cc}
				 e^{- \imath \omega_i} &  0  \\
				 0 & e^{\imath \omega_i} \\
			\end{array} \right)
			\left( \begin{array}{cc}
				 \tau_{i,1} & \imath \sqrt{1-\tau_{i,1}^2}  \\
				 \imath \sqrt{1-\tau_{i,1}^2} & \tau_{i,1}  \\
			\end{array} \right)			
			\left( \begin{array}{cc} 
				 1 &  0  \\
				 0 & e^{\imath \phi_i} \\
			\end{array} \right)
		\end{equation}
	\end{widetext}
	\item introduce gaussian noise on the four parameters $(\tau_{i,1},\tau_{i,2},\omega_i,\phi_i)$, by sampling new values from a normal distribution centered on the ideal ones and with widths $\sigma_{BS}$ and $\sigma_{PS}$ respectively for ($\tau_{1},\tau_{2}$) and for ($\omega,\phi$);
\item generate new unit cells from the noisy values, adding possible losses $\textup{diag}(\eta_i)$ at the output, and rebuild the noisy $U$.
\end{enumerate}

\begin{figure*}[t!]
\centering
\includegraphics[width=0.95\textwidth]{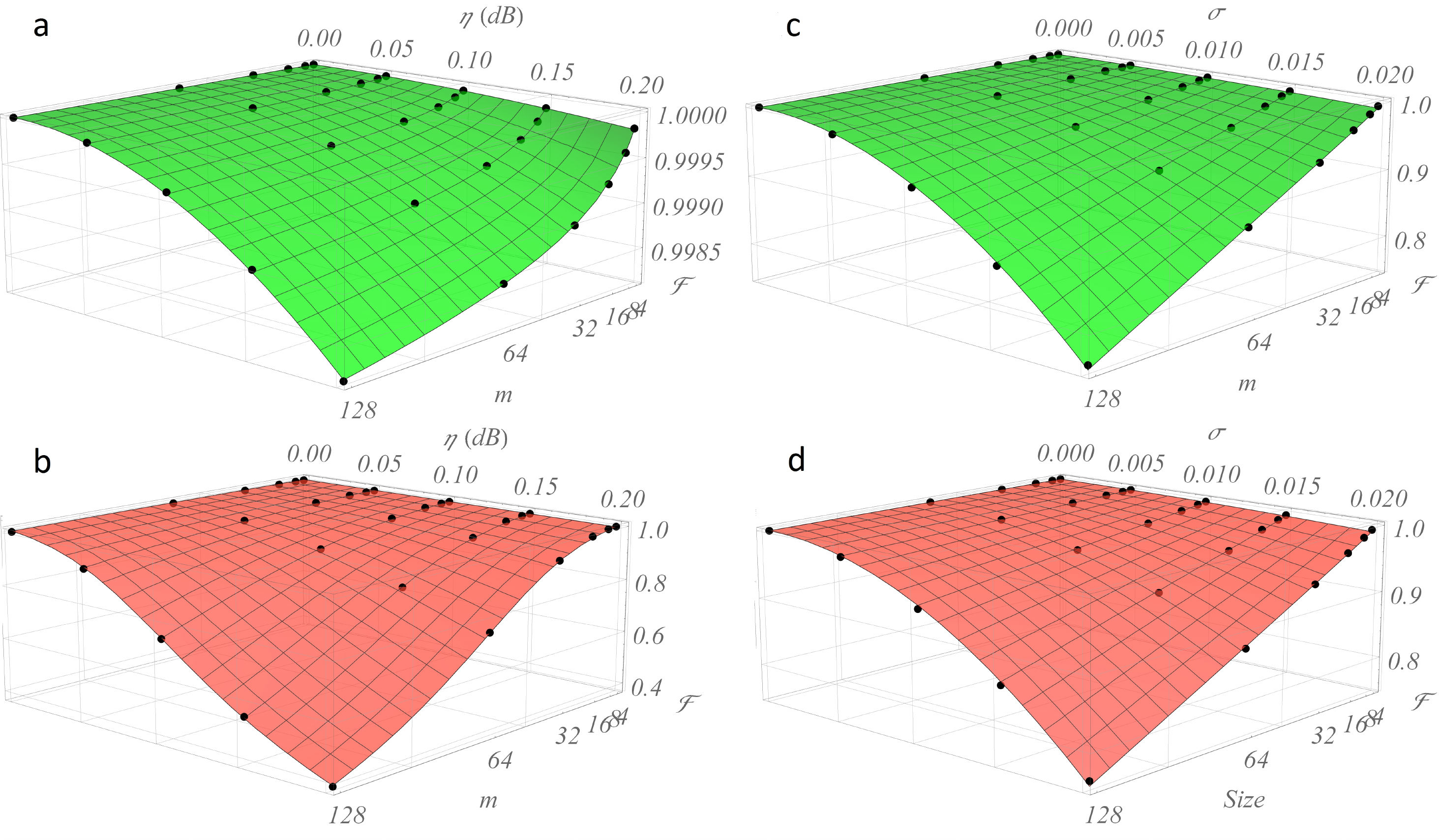}
\caption{{\bf Haar-random with losses and noise}. Noise and losses affect the implementation of Haar-random unitaries in different ways in the $C$- and $R$-designs. {\bf a-b}, average fidelity $\textup{F}$ for different levels of loss $\eta$ per beam splitter and size $m$ of the interferometer in the $C$ ({\bf a}) and $R$ ({\bf b}) designs. Note the difference in vertical scale, due to the more balanced structure of $C$ where the dependency on $\eta$ arises from the slight path asymmetry of the outer waveguides. {\bf c-d}, average fidelity $F$ for different levels of noise $\sigma$ in the optical elements and size $m$, averaged over $500$ noisy unitaries. Here noise is treated equally on both beam splitters and phase shifters, namely $\sigma=\sigma_{BS}=\sigma_{PS}$. Note that the scaling is identical in ({\bf c}) and  ({\bf d}) since, averaging over the unitaries, the deterioration of $F$ is due to the number of noisy elements, which is the same in the two schemes. Surfaces: heuristic non-linear fits of the data (see Supplementary Note 1 online).}
\label{fig:Losses+NoiseRvsC}
\end{figure*}

This simple procedure allows us to investigate the effect of noise and losses on the $C$- and $R$-designs for the implementation of Haar random unitaries. Incidentally, our numerical simulations confirm \cite{Russell17} that predictions would be remarkably different if, instead of generating every time a new unitary according to the Haar measure, we directly generated sets of uniformly distributed random parameters $(\omega_i,\phi_i)$ to -mistakenly- speed up the calculation. Subsequently, the same analysis is repeated for all three architectures ($C$, $R$ and $F$) focusing on the implementation of the Fourier and the Sylvester transformation.

\begin{figure*}[t!]
\centering
\includegraphics[width=0.99\textwidth]{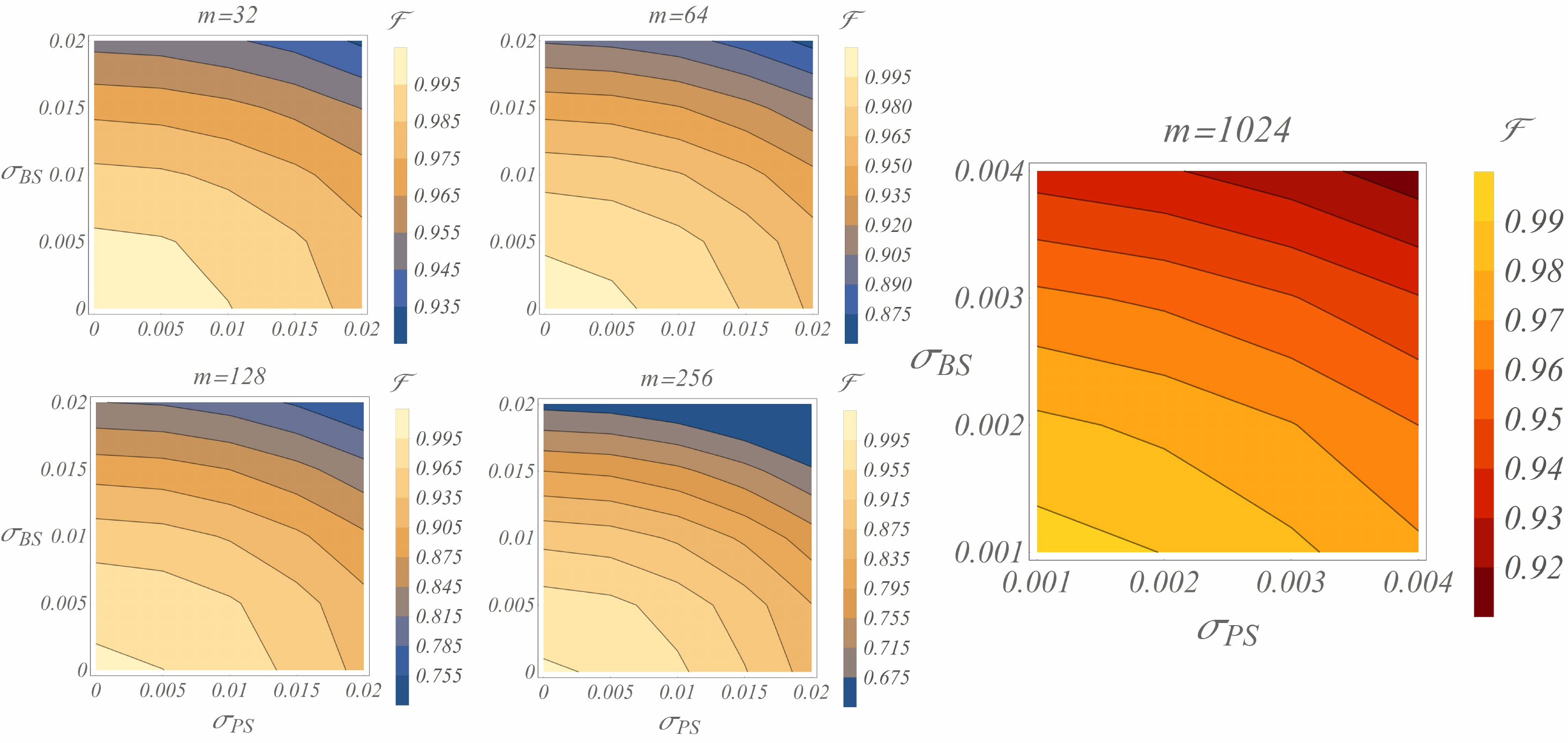}
\caption{{\bf Haar-random with non-ideal transmissivities and phase shifts}. Fabrication imperfections and errors in real-time control over the phases are in general on different scales, depending for instance if we are considering (un)balanced beam splitters or (not-)reconfigurable phase shifters. Here noisy Haar-random implementations are investigated under different values of $\sigma_{BS}$ and $\sigma_{PS}$, to analyze the separate contribution of each source of error to the final unitary transformation. Since the average robustness against noise of the $C$- and $R$-designs is identical (see Fig.\ref{fig:Losses+NoiseRvsC}), here the contour plots are retrieved via $R$ decompositions for $m=32,64,128,256$, while an optimized version of $C$ decomposition is employed for $m=1024$ (see Supplementary Note 2). }
\label{fig:HaarSeparateNoise}
\end{figure*}

\subsection*{Benchmarking Haar-random interferometers} 

The results of our first analyses are shown in Fig.\ref{fig:Losses+NoiseRvsC}. The figure of merit adopted to compare the two architectures is the fidelity \cite{Clements16} $\mathcal{F}=| \frac{\textup{Tr}(U_{imp}^\dagger \, U)}{\sqrt{m \, \textup{Tr}(U_{imp}^\dagger \, U_{imp})}} | ^2 $, being $U$ a $m \times m$ unitary transformation and $U_{imp}$ its imperfect implementation. Thanks to the rescaling factor in the denominator, accounting for the contribution of lossy implementations, this fidelity is particularly suitable to characterize non-unitary transformations.

Fig.\ref{fig:Losses+NoiseRvsC}a,b show the deterioration of $\mathcal{F}$ for increasing values of losses $\eta$ per unit cell and size $m$ of the circuit for the $C$- and $R$- designs. In this simulation, optical elements are assumed to be immune to fabrication noises in order to isolate the $\eta$-contribution. As pointed out in Ref.\cite{Clements16}, the $C$-design is more resilient to internal losses than the $R$- design, thanks to the almost total balance between the mode interactions. Moreover, as the heuristic non-linear fit suggests (see Supplementary Note 1), also the scaling of the fidelity is much more favorable for the former architecture, features that pushes $C$ as a promising candidate for large-scale platforms or where, due to technical issues inherent to the specific implementation, it is not possible to guarantee a low level of losses in each optical element. Fig.\ref{fig:Losses+NoiseRvsC}c,d show instead the average effect of a noisy implementation of the optical components over the fidelity.  Similarly to the previous analysis, our simulation is carried out for various levels of noise $\sigma$ and different sizes $m$, aiming to retrieve a more complete feeling of its scaling. Noise is assumed to be of equal intensity at this stage on both beam splitters transmissivities ($\sigma_{BS}$) and phase shifts ($\sigma_{PS}$), i.e. $\sigma=\sigma_{BS}=\sigma_{PS}$, in order to capture the scaling of the performance in a unique three-dimensional plot. Our simulations confirm previous qualitative estimates \cite{Clements16} concerning the similarity of the scalings of the average fidelity in the two architectures, providing a clear and quantitative picture of its dependency on the fabrication noise over the optical elements.

\begin{figure*}[t!]
\centering
\includegraphics[width=0.99\textwidth]{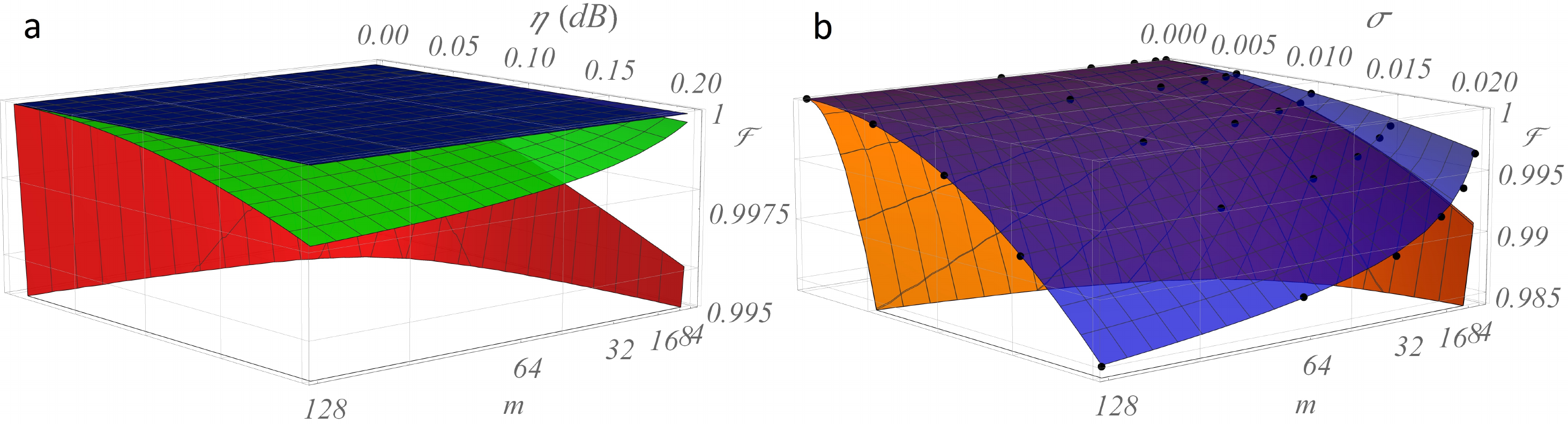}
\caption{{\bf Generalized Hadamard transformations with $C$-, $R$- and $F$ designs}. While universal for unitary decompositions, the $C$- and $R$- architectures offer suboptimal solutions for the implementation of specific classes with higher symmetries. The Fast architecture is optimized for the implementation of the quantum Fourier transform and the class of Hadamard transforms. {\bf a}, Average fidelity as a function of losses per unit cell and network size  (blue, green and red surfaces for $F$-, $C$- and $R$-designs respectively). Full symmetry between the optical paths cancels out the effect of constant losses per unit cell in the $F$- design. {\bf b}, Fast architectures (blue surface) are also more resilient to fabrication noise, thanks to the reduced depth of the circuit. Here, only one (orange) surface is shown for both $C$- and $R$-designs, assuming equal resilience to noise. Surfaces: heuristic non-linear fits of the data (see Supplementary Note 1 online).}
\label{fig:FastVsReckClements}
\end{figure*}

\begin{figure}[hb!]
\centering
\includegraphics[width=0.48\textwidth]{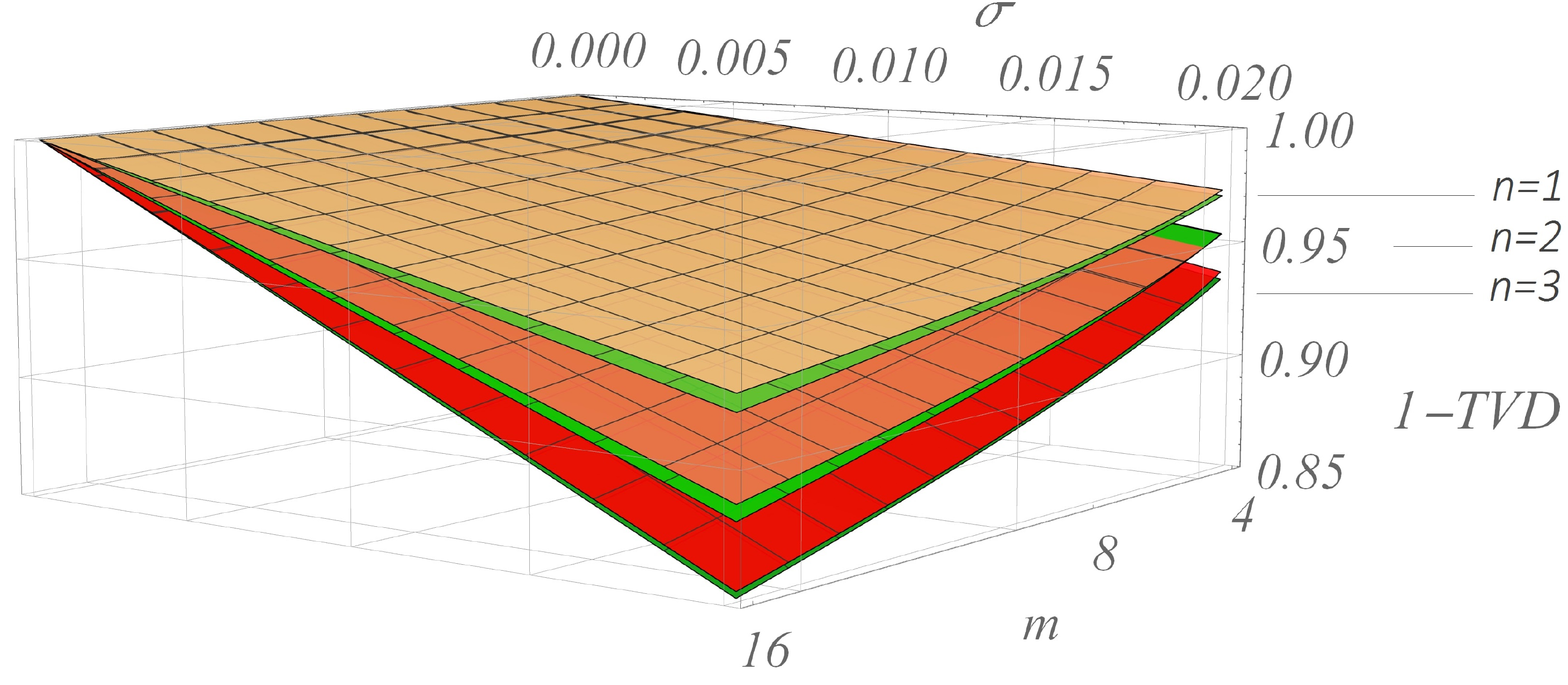}
\caption{{\bf Multiphoton interference in noisy Haar-random interferometers}. Several applications require to evolve many-photon states in large-scale quantum circuits. The total variation distance (TVD) is in this sense a good estimate of the goodness of an experimental implementation, being it a natural measure of the distance between two probability distributions. Here, plots compare the ideal and noisy output probability distributions relative to $n=1,2,3$-photon collision-free states, averaged over all the inputs. For each $n$, $100$ unitaries are sampled and implemented according to the $C$- (green) and $R$- (red) designs setting $\sigma_{BS}=\sigma_{PS}=\sigma$. Our analysis confirms the prediction of Fig.\ref{fig:Losses+NoiseRvsC}, where the $R$-design is found to be slightly more robust againt noise, though this difference seems to become negligible for increasing values of $n$. Surfaces: heuristic non-linear fits of the data (see Supplementary Note 1 online).}
\label{fig:TVDRvsC}
\end{figure}

After this preliminary stage, more general investigations have been carried out by considering different levels of noise over the fabrication transmissivities and phases, i.e. studying the practical situation where $\sigma_{BS}\ne \sigma_{PS}$. Results of this analysis are shown in the contour plots of Fig.\ref{fig:HaarSeparateNoise} for different sizes of the interferometers, highlighting a number of interesting features. First, the qualitative dependency on the noise seems to remain fixed while increasing the dimension of the circuit, though the average fidelity rapidly drops to low values already at $m=64$, for intensities of noise that meet the criteria of the technological state of the art. Looking at future large-scale implementations, we see that for $m=1024$ even a very low level of noise makes the average fidelity drop to values as low as $\sim 0.90$. Note the shift in the axes scales between the four left contour plots ($m\le256$) and the one on the right ($m=1024$). Moreover, we observe that the two sources of noise affect the fidelity in a similar way, with an intensity approximately double for the one on the transmissivities in the Mach-Zehnder. Thus, while dynamic control over the phases can in principle mitigate the effects of noisy implementations, non-ideal values of the transmissivities of the symmetric beam splitters remain critical when looking at large-scale implementations.

So far, we adopted as figure of merit the fidelity of the quantum process. Though perfectly suitable to benchmark noisy transformations, this quantity fails in general to highlight more complex many-particle phenomena. This requirement becomes even more strict for instance in the context of Boson Sampling and, more specifically, in its validation, where multiphoton interference plays a key role to guarantee the computational complexity of the problem. Indeed, while Boson Sampling preserves its complexity even in lossy and imperfect conditions below a certain threshold \cite{Arkhipov14,Leverrier15}, multiphoton interference in Hadamard interferometers \cite{Crespi2016,Viggianiello17} was shown to be a promising tool to correctly validate its operation. Partial deviation from their ideal symmetric structures can then spoil the interference effects in the output distributions. For this reason, we investigated the performance of noisy architectures also in the scope of multiphoton output probability distributions, employing as figure of merit the total variation distance (TVD) between the ideal and the actual $n$-photon distributions. Results for this analysis are shown in Fig.\ref{fig:TVDRvsC}: again, noise affects almost equally the $C$ and $R$ architectures when averaging over all the input/output states. The slight difference in favor of the $R$-design may not be practically appreciable in real experimental conditions and it rapidly becomes negligible for higher values of ($n$,$m$). Thus, the two architectures behave equally as far as lossless multiphoton investigations are concerned.

\begin{table*}[!htbp]
\renewcommand{\arraystretch}{1.55}
\begin{tabular}{*{10}{c}}
\hline
 &  \multicolumn{9}{p{9cm}} {\centering Quantum Fourier transform and Sylvester} \\ \hline

 & \multicolumn{3}{p{3cm}}{\centering $m=64$} & \multicolumn{3}{|p{3cm}|}{\centering $m=128$} & \multicolumn{3}{p{3cm}}{\centering $m=256$} \\ \cline{2-10}

\hfill \; $\sigma_{BS} \; \rightarrow$ & $0.005$ & $0.01$ & $0.02$ & $0.005$ & $0.01$ & $0.02$ & $0.005$ & $0.01$ & $0.02$ \\ \hline

 \multirow{ 2}{*}{$\sigma_{PS}= 0.001$\;} & 0.999 & 0.995 & 0.981 & 0.998 & 0.994 & 0.976 & 0.998 & 0.992 & 0.972 \\    & 0.994 & 0.975 & 0.904 & 0.987 & 0.950 & 0.817 & 0.975 & 0.903 & 0.667 \\ \hline

 \multirow{ 2}{*}{$\sigma_{PS}= 0.01$\;} & 0.998 & 0.994 & 0.980 & 0.997 & 0.993 & 0.975 & 0.997 & 0.992 & 0.971 \\    & 0.984 & 0.966 & 0.895 & 0.969 & 0.933 & 0.803 & 0.938 & 0.870 & 0.644 \\ \hline

 \multirow{ 2}{*}{$\sigma_{PS}= 0.02$\;} & 0.995 & 0.991 & 0.976 & 0.994 & 0.989 & 0.971 & 0.993 & 0.988 & 0.968 \\    & 0.956 & 0.939 & 0.870 & 0.917 & 0.882 & 0.759 & 0.838  & 0.770 & 0.575 \\ \hline

\end{tabular}
\caption{{\bf Benchmarking Hadamard transformations}. Generalized Hadamard transformations benefit from optimized architectures \cite{Cooley65,Barak07}. Here, average fidelities (in each row, $F$: up; average of $C$ and $R$: down) are reported for each combination of size $m$ and noises $\sigma_{BS}, \sigma_{PS}$, averaging over $100$ noisy unitaries. Data for the Fourier and Sylvester transformations are displayed in a single table since the values in the two cases are equal within a discrepancy lower than $\sim 0.001$.}
\label{tbl:QFTandS}
\end{table*}

We conclude from our analysis that the $C$- and $R$-designs are ultimately equivalent in terms of resilience to noise when averaging over all input/output configurations. On one side, the single optical elements affect in a different way the elements of the unitary transformation in the two schemes, being the $C$ and $R$ parameters more localized respectively in the upper corner and central part of the unitary matrix. However, for practical noise levels and general multi-input applications, this difference does not give rise to any effective deviation between the two schemes. In contrast, the two architectures behave differently when it comes to lossy implementations, where the symmetric $C$-design outperforms the $R$ scheme.

\subsection*{Benchmarking Fourier and Sylvester interferometers}

Though the $C$- and $R$-designs are universal for unitary evolutions, often it is desirable to have circuits optimized for specific relevant tasks. It is the case of the quantum Fourier or Sylvester transforms, which have importance on their own in different scopes of quantum information processing. 
In this section we investigate the performance of the Fast ($F$) scheme \cite{Barak07,Crespi2016, Flamini16, Viggianiello17} for the implementation of generalized Hadamard transformations \cite{Flamini17} and compare it with the average performance of the two universal designs, following the same procedure outlined in Fig.\ref{fig:analysis}.
Fig.\ref{fig:FastVsReckClements} reproduces the analysis of Fig.\ref{fig:Losses+NoiseRvsC} including the $F$ architecture. Being the circuit completely symmetric, the structure is totally immune to constant propagation losses, beating even the highly resilient $C$-design. The $F$-design is also more resilient to noise, thanks to the reduced depth of the circuit which lowers the number of noisy optical elements from $\frac{m}{2} (m-1)$ to $\frac{m}{2} \log{m}$.
A benchmark summary of this comparison is reported in Table \ref{tbl:QFTandS} for both the quantum Fourier transform and the Sylvester interferometers.

\subsection*{Conclusions}

Photonic technologies promise to enable the application of several quantum information protocols, ranging from fundamental research to quantum computation and optical quantum networks. In this work we have provided a comprehensive analysis of the performance of the three main interferometric schemes, namely the triangular \cite{Reck} and square \cite{Clements16} designs and the Fast architecture, under realistic conditions of losses and noise. Our investigations quantitatively address the issue of imperfect implementations for interferometers of increasing dimension, aiming to embrace both mid-term and long-term technological standards. Our results confirm the qualitative expectation that the square design performs way better, in terms of fidelity, between ideal and lossy evolutions with respect to the triangular one, even though the latter exhibits a slightly enhanced resilience to fabrication noise. Thus, we conclude that the square design is preferable in practical applications involving multi-input protocols and Haar-random generations, especially in high-dimension circuits where the issue of propagation losses becomes critical.

Fast architectures represent instead a specialized design to implement a significant class of unitary evolutions, highly optimized for the realization of Fourier \cite{Crespi2016,Flamini16} and Sylvester \cite{Viggianiello17} quantum transformations. Our results quantitatively highlight the improved performances of this scheme with respect to the universal ones, thus making it the preferred choice when applicable. Furthermore, we have provided a closed-form expression for the elements of the unitary describing these circuits, mapping them to the transmissivities and the phase shifts in the real optical implementation. Due to the deep relevance of this class of quantum routines, and thanks to the high enhancements offered by their optimized implementations, we expect Fast architectures to gain a key role among photonic platforms in synergy with the universal schemes, to fully benefit from the unique advantages of both designs.

\vspace{1em}
During the completion of this manuscript, related work has been reported in \cite{Russell17,Burgwal17}.

\subsection*{Acknowledgements} 
This work was supported by the ERC-Starting Grant 3D-QUEST (3D-Quantum Integrated Optical Simulation; grant agreement no.307783): http://www.3dquest.eu, and by the H2020-FETPROACT-2014 Grant QUCHIP (Quantum Simulation on a Photonic Chip; grant agreement no. 641039): http://www.quchip.eu.

%


\end{document}


\title{Supplementary Information - Benchmarking integrated photonic architectures}

\author{Fulvio Flamini}
\affiliation{Dipartimento di Fisica, Sapienza Universit\`{a} di Roma, Piazzale Aldo Moro 5, I-00185 Roma, Italy}

\author{Nicol\`o Spagnolo}
\affiliation{Dipartimento di Fisica, Sapienza Universit\`{a} di Roma, Piazzale Aldo Moro 5, I-00185 Roma, Italy}

\author{Niko Viggianiello}
\affiliation{Dipartimento di Fisica, Sapienza Universit\`{a} di Roma, Piazzale Aldo Moro 5, I-00185 Roma, Italy}

\author{Andrea Crespi}
\affiliation{Istituto di Fotonica e Nanotecnologie, Consiglio Nazionale delle Ricerche (IFN-CNR), Piazza Leonardo da Vinci, 32,
I-20133 Milano, Italy}
\affiliation{Dipartimento di Fisica, Politecnico di Milano, Piazza
Leonardo da Vinci, 32, I-20133 Milano, Italy}

\author{Roberto Osellame}
\affiliation{Istituto di Fotonica e Nanotecnologie, Consiglio Nazionale delle Ricerche (IFN-CNR), Piazza Leonardo da Vinci, 32,
I-20133 Milano, Italy}
\affiliation{Dipartimento di Fisica, Politecnico di Milano, Piazza
Leonardo da Vinci, 32, I-20133 Milano, Italy}

\author{Fabio Sciarrino}
\affiliation{Dipartimento di Fisica, Sapienza Universit\`{a} di Roma,
Piazzale Aldo Moro 5, I-00185 Roma, Italy}

\maketitle

\section*{Supplementary Note 1: Details on the non-linear fits}
\label{.....}

Below we provide details on the non-linear heuristic fits shown in the main text. Functions are chosen by selecting the dependency that best matched the simulated data points in the region under investigation, while keeping low the number of free parameters to avoid overfitting. The Fidelity $\mathcal{F}$ as a function of the size of the circuit $m$ and loss per beam splitter $\eta$ (in dB) has the following expressions, respectively for the $C$- [S1] and $R$-designs [S2]

\begin{equation}
\mathcal{F}_C (m,\eta) \sim 1- A_c\; \eta^2\, \log(B_c \, m + C_c)  \qquad \quad A_c=0.0158 \pm 0.0002, \quad B_c=0.140 \pm 0.005, \quad C_c=0.79 \pm 0.01
\end{equation}
\begin{equation}
\mathcal{F}_R (m,\eta) \sim  \frac{A_r+e^{ B_r \, m \, \eta}}{A_r+e^{ C_r \, m \, \eta}} \quad \qquad A_r=7.55 \pm 0.51, \quad B_r=0.163 \pm 0.007, \quad C_r=0.204 \pm 0.007
\end{equation}
as retrieved from a non-linear fit in the region $m \in [4,256]$ and $\eta \in [0,0.2]$, by sampling $500$ Haar-random lossy unitaries for each simulated point.
Similarly, $\mathcal{F}$ as a function of $m$ and noise $\sigma$ has the expressions ($F$: Fast scheme)

\begin{equation}
\mathcal{F}_C (m,\sigma) \sim 1-\alpha \; m \; \sigma^2 \qquad \qquad \alpha=4.932 \pm 0.034
\end{equation}
\begin{equation}
\mathcal{F}_R (m,\sigma) \sim 1-\beta \; m \; \sigma^2 \qquad \qquad \beta=4.896 \pm 0.036
\end{equation}
\begin{equation}
\mathcal{F}_F (m=2^n,\sigma) \sim 1- \gamma \, n \,\sigma^2 \qquad \qquad \gamma=7.73 \pm 0.04
\end{equation}

as retrieved from a non-linear fit in the region $m \in [4,128]$ and $\sigma \in [0,0.02]$, by sampling $500$ Haar-random noisy unitaries for each simulated point. We observe that the effective dependency of $\mathcal{F}$ is indeed on the depth of the circuit, i.e. $O(m)$ for $C$ and $R$, $O(\log m)$ for $F$.
Finally, the 3-photon total variation distance as a function of $m$ and $\sigma$ has the expression

\begin{equation}
TVD_{C,R}^{(3)} (m,\sigma) \sim 1-A_c \;\sigma \, \sqrt{m(m + B_c )}  \qquad \qquad A_c=0.23 \pm 0.01, \quad B_c=47 \pm 5
\end{equation}

which is found to be approximately the same for both $C$ and $R$ schemes. The formula is estimated in the region $m \in [4,16]$ and $\sigma \in [0,0.02]$, by sampling $100$ Haar-random noisy unitaries for each simulated point.
\vspace{5em}

\section*{Supplementary Note 2: Square decomposition with high-dimensional unitaries}
\label{...}

Below we provide a sketch of the routine adopted for the $C$ decomposition of high-dimensional unitaries [S1]. Our routine is formally equivalent to the original algorithm but it avoids matrix multiplications, thus significantly decreasing the computational resources required for large interferometers. Steps marked with (*) can be implemented easily without a full matrix multiplication, since each $T$ matrix affects only one pair of rows/columns of $U$.

\vspace{5em}

\noindent First, following [S1], we retrieve the $\frac{m(m-1)}{2} $ pairs of parameters $ \{(\phi,\omega) \} \cup \{ (\phi^\dagger,\omega^\dagger)\}$ \newline
\vspace{2em} 

\qquad For i = 1 ... m \newline

\vspace{-0.75em}
\qquad \qquad if i == odd: \qquad For j = 0 ... i - 1 \newline

\vspace{-0.25em}
\qquad \qquad \qquad \qquad \qquad \qquad \qquad$\phi_s^\dagger =  \textup{arg} (U_{m-j,i-j}) - \textup{arg} (U_{m-j,i-j+1}) $  \newline

\vspace{-0.75em}
\qquad \qquad \qquad \qquad \qquad \qquad \qquad$\omega_s^\dagger = \textup{arctg} \left( \frac{U_{m-j,i-j+1}}{U_{m-j,i-j}} \, e^{\i \phi_s^\dagger} \right) $ \newline

\vspace{-0.75em}
\qquad \qquad \qquad \qquad \qquad \qquad \qquad$ U \rightarrow U \, . \, T_{i-j,i-j+1}^\dagger \left( \phi_s^\dagger,\omega_s^\dagger \right) $ \qquad \qquad \qquad \; (*) \newline

\vspace{-0.25em}
\qquad \qquad else: \qquad \qquad \; For j = 0 ... i - 1 \newline

\vspace{-0.25em}
\qquad \qquad \qquad \qquad \qquad \qquad \qquad$\phi_s = \textup{arg} (U_{m+j-i+1,j+1}) -  \textup{arg} (U_{m+j-i,j+1}) $  \newline

\vspace{-0.75em}
\qquad \qquad \qquad \qquad \qquad \qquad \qquad$\omega_s = \textup{arctg} \left( \frac{U_{m+j-i,j+1}}{U_{m+j-i+1,j+1}} \, e^{\i \phi_s} \right) $ \newline

\vspace{-0.75em}
\qquad \qquad \qquad \qquad \qquad \qquad \qquad$ U \rightarrow T_{m+j-i,m+j-i+1} \left( \phi_s,\omega_s \right)  . \,  U $ \qquad \qquad (*)  \newline

\vspace{2em} 
\noindent Then we need to move the remaining diagonal matrix $D$ to the left of the decomposition [S1], which leads us to a new set of parameters $ \{ (\phi,\omega)\} $. By defining $\delta = \{ \textup{arg} \left( \textup{diag} \, D \right) \} $ and $\chi$ as the $\left( \frac{m (m-1)}{2} \times 2 \right)$-dimensional list of pairs of modes mixed step by step by the $N=\sum_{i=0}^{\ceil*{\frac{m}{2}}-1}(2 i)$ matrices $T$ (as given by the order of the decomposition), we iteratively update the $ \{ \phi\} $ as \newline

\vspace{2em} 
\qquad For $s = 1 \, ...\, N $ \newline

\vspace{-0.75em}
\qquad \qquad $ tmp = \delta_{\chi_{s,2}} $ \newline

\vspace{-0.75em}
\qquad \qquad $ \delta_{\chi_{s,2}} =  \delta_{\chi_{s,1}} + \phi_s $ \newline

\vspace{-0.75em}
\qquad \qquad $ \phi_s =  \delta_{\chi_{s,1}} - tmp $ \newline

\vspace{2em} 
\noindent The final list of parameters $\pi$ is then \qquad $\pi=\{(\phi,\omega) \} \cup \textup{Reversed} \{ (\phi^\dagger,\omega^\dagger)\} $, \\ \noindent where \textit{Reversed} takes the list of pairs of parameters $\{ (\phi^\dagger_s,\omega^\dagger_s)\}_{s=1...n} $ in the reversed order: $\{ (\phi^\dagger_s,\omega^\dagger_s)\}_{s=n...1} $.

\vspace{5em}

\section*{Supplementary References}

\noindent [S1] W. R. Clements, P. C. Humphreys, B. J. Metcalf, W. S. Kolthammer, and I. A. Walmsley, Optica {\bf 3}, 1460-1465 (2016).

\noindent [S2] M. Reck, A. Zeilinger, H. J. Bernstein, and P. Bertani, Phys. Rev. Lett. {\bf 73}, 58 (1994).